\documentclass[aip,preprint]{revtex4-2}
\usepackage{graphicx}
\usepackage{amsmath}
\usepackage{xcolor}
\usepackage{adjustbox}
\usepackage{chemformula}
\usepackage{multirow}
\usepackage{array}
\usepackage{dcolumn}
\usepackage{bm}
\usepackage[mathlines]{lineno}
\usepackage{subcaption}
\usepackage[normalem]{ulem}
\setcitestyle{maxcitenames=2}
\draft 

\begin{document}
\title{Compressibility effects on the flow-induced motions of bluff bodies: common features from transverse and rotary oscillations}
\author{Sawant Omkar Deepak }
\email{omkars30.20dr0128@mech.iitism.ac.in}
\affiliation{Department of Mechanical Engineering, IIT(ISM) Dhanbad, India-826004}
\author{Swagata Bhaumik }
\email{swagatabhaumik@mech.iitism.ac.in}
\affiliation{Department of Mechanical Engineering, IIT(ISM) Dhanbad, India-826004}
\date{\today}

\begin{abstract}


Compressibility effects on flow-induced motions of bluff bodies are examined through two-dimensional direct numerical simulations of undamped transverse vortex-induced vibration (TVIV) of a circular cylinder ($Re=150$) and vortex-induced rotary oscillation (VIRO) of an elliptic cylinder of aspect ratio $0.5$ ($Re=200$) for free-stream Mach numbers $0.1\leq M_{\infty}\leq 0.6$. Response characteristics are analyzed across the first desynchronized zone ($DS\text{-}I$), the synchronous (lock-in) branch ($SB$), and the second desynchronized zone ($DS\text{-}II$). While compressibility effects are noted to stabilize the oscillations in $SB$ in terms of reductions in response amplitudes and fluctuating aerodynamic coefficients, as well as the extent and intensity of hysteretic jumps, in contrast, the oscillation amplitudes and aerodynamic coefficients increase with $M_{\infty}$ in $DS\text{-}II$, where the amplitude variation with $M_{\infty}$ is shown to follow the \textit{modified Prandtl-Glauert relationship} for both TVIV and VIRO cases. This indicates the role of the inviscid nature of compressibility in the $DS\text{-}II$ regime in increasing the oscillation response, as opposed to $SB$ where enhanced viscous stabilization with increasing $M_{\infty}$ leads to the moderation of response. Although the simulations are performed at low Reynolds numbers, the identified trends are consistent with fundamental compressibility mechanisms and may therefore persist at higher Reynolds numbers.

\end{abstract}
\pacs{}
\keywords{Compressible flow; Flow induced oscillation; Prandtl-Glauert relationship; Flow past bluff bodies; Direct numerical simulation; Computational fluid dynamics}
\maketitle

\section{Introduction}\label{sec1}

Flow-induced vibration (FIV) arises from the interaction of unsteady aerodynamic forces with a flexible structure and is encountered across a wide range of engineering systems. While most classical studies have focused on incompressible, low-speed regimes, modern aerospace applications—such as control surfaces, guided missiles, and launch vehicles—operate in the compressible regime, where density variations can significantly alter the phase relationship between aerodynamic loading and structural motion. These effects can lead to phenomena such as transonic flutter, shock buffeting, and acoustic fatigue. The present work examines two representative classes of flow-induced motion — transverse vibration of a circular cylinder and rotary oscillation of an elliptic cylinder — in the low-to-moderate subsonic range to identify and characterize common features of compressibility effects. The simulations are performed at low Reynolds numbers so that two-dimensional flow remains physically meaningful while allowing extensive parametric variation. Although the use of low $Re$ at higher $M_{\infty}$ is not physically realizable, it enables a clean numerical isolation of compressibility effects on the response branches.

Over the past two decades, numerical investigations have substantially advanced the understanding of FIV mostly for incompressible flows. \citet{singh2005vortex} examined the low-Reynolds-number response of a freely vibrating circular cylinder and demonstrated the occurrence of hysteresis and distinct vortex-shedding modes, while \citet{prasanth2008vortex} subsequently provided a detailed characterization of transverse VIV in the laminar regime, including synchronization, response characteristics, and wake transitions. \citet{leontini2006beginning} further identified distinct response branches through two-dimensional simulations, establishing that branch-like behavior can persist even in the low-Reynolds-number regime. Numerical investigations of non-circular bluff bodies have elucidated the intricate coupling among body geometry, unsteady fluid loading, wake dynamics, and structural motion (vibration amplitude, synchronization characteristics, response branches, hysteresis, and transitions between distinct FIV regimes, \textit{e.g.}, VIV, desynchronized response, and galloping) that underpins the incompressible FIV response \cite{sen2011free,zhao2013numerical,zhu2021flow,navrose2014free,kumar2018modification}. Such studies have also been extended to incorporate the effects of forced, free, and mixed convection on FIV of bluff bodies \cite{biswas2009effect,ding2024exploring,Sohankar2018,wan2016suppression,garg2019vortex,wan2021vortex}, demonstrating that Richardson number, flow orientation, thermal conditions, localized heating, and cylinder cross-section strongly affect vortex shedding, VIV response, and resonance characteristics, along with the suppression of galloping behavior.

Compressibility has been shown to modify the wake topology and vortex-shedding characteristics of stationary circular cylinders \cite{canuto2015two,nagata2020experimental}, providing a basis for examining its role in freely vibrating systems. Despite the difference in the $Re$ range investigated, both authors noted that an increase in $M_{\infty}$ tends to suppress fluctuations. \citet{song2024generalized} investigated undamped transverse VIV of a circular cylinder at $Re=100$ and concluded that the compressibility effects start to affect the vibration characteristics for free-stream Mach numbers ($M_{\infty}$) greater than $0.1$ by comparing the vibration characteristics at a single reduced velocity within the lock-in zone. A similar investigation by \citet{rajamuni2024immersed} for $Re=200$ and $M_{\infty}=0.2$ showed that transverse VIV substantially intensifies acoustic radiation through the enhanced unsteady fluid--structure interaction. Results presented in \citet{liu2025fluid} for a freely vibrating circular cylinder with non-linear elastic support at $Re=150$ and $M_{\infty}=0.2$ showed that structural non-linearity substantially modifies the response branches, lock-in characteristics, and transitions between synchronized and desynchronized states along with associated acoustic radiation. \citet{wang2025aerodynamic} further characterized the aerodynamic and acoustic responses of a transversely vibrating circular cylinder at $Re=150$ and $M_{\infty}=0.2$, identifying different response regimes, including lock-in, phase-switching, and beating behavior and their subsequent influence on the corresponding acoustic radiation. However, a systematic investigation of compressibility effects on the flow-induced motion of bluff bodies across different response regions is not reported in the literature. This is the motivation of the present investigation: to identify and characterize the common features of compressibility on flow-induced oscillation of bluff bodies and extract general, distinct patterns associated with it. We consider both transverse vortex-induced vibration (TVIV) of a circular cylinder and vortex-induced rotary oscillation (VIRO) of an elliptic cylinder of aspect ratio $0.5$ at $Re=150$ and $200$, respectively, while $M_{\infty}$ is varied from $0.1$ to $0.6$.

This article is organized as follows. \textbf{Section~\ref{sec2}} includes a detailed discussion of the problem statement, the grid generation strategy employed, the computational methods used, the governing equations, and boundary conditions. Results and discussion are presented in \textbf{Section~\ref{sec3}}, while \textbf{Section~\ref{sec4}} reports the summary and conclusions.

\section{Governing equations, numerical methods, and boundary conditions} \label{sec2}

Direct numerical simulation (DNS) of $2D$ unsteady, viscous, compressible flow is performed for both undamped transverse VIV (TVIV) of a circular cylinder and undamped vortex-induced rotary oscillation (VIRO) of an elliptic cylinder using an in-house developed high-accuracy compressible Navier-Stokes solver. Here, the fluid is assumed to be ideal and calorically perfect with identical parameters as given in \citet{deepak2026}. The temperature-dependent variation of dynamic viscosity is assumed to follow Sutherland's law \cite{hilsenrath1955}. The Reynolds number $Re$ is defined with respect to $D$ (diameter of the circular cylinder and major axis of the elliptic cylinder), free-stream velocity $U_{\infty}$, free-stream density $\rho_{\infty}$, and dynamic viscosity $\mu_{\infty}$, while $M_{\infty}= {U_{\infty}}/{a_{\infty}}$, where $a_{\infty}$ represents the speed of sound in the free-stream.

The non-dimensional structural dynamic equations governing the time variation of undamped oscillation for the TVIV case and the VIRO case, respectively, are given as
\begin{eqnarray}
  \ddot{y}+\left(\dfrac{2\pi}{U^*}\right)^2y&=&\dfrac{2}{\pi m_*}C_l \\
  \ddot{\theta} + \left( \frac{2\pi}{U^*} \right)^2 {\theta} &=& \frac{1}{2} \left(\frac{1}{m_* A R_{g}^2} \right) C_m
\end{eqnarray}
\noindent Here, the undamped oscillation for both the elliptic and circular cylinders is governed by the reduced velocity
\begin{equation}
  U^*= \frac{U_{\infty}}{f_n D}
\end{equation}
\noindent and the reduced mass $m_*$ (ratio of average material density to $\rho_{\infty}$), where $f_n$ indicates the natural frequency of structural oscillation; $C_{l}$ is the coefficient of lift; $A=\pi A_r/4$ denotes the non-dimensional cross-sectional area of the elliptic cylinder having aspect ratio $A_r$; $C_{m}$ is the coefficient of aerodynamic moment; and $R_{g}=\sqrt{1+A_r^2}/(2\sqrt{2})$ represents the non-dimensional radius of gyration of the elliptic section about the center. For TVIV of a circular cylinder, $m_*=3.183$, while for VIRO of an elliptic cylinder, $A_r=0.5$ and $m_*=6.366$ are used here. It may be noted that for TVIV, $f_n=\sqrt{k_y/m}/(2\pi)$, while for VIRO, $f_n=\sqrt{k_t/I_z}/(2\pi)$, where $k_y$ and $k_t$ are the spring constants for transverse vibration and rotary oscillation motions, respectively, $m$ is the mass of the cylinder, and $I_z$ indicates the rotary moment of inertia of the cylinder. The parameters $U^*$ and $m_*$ are non-dimensional quantities, such that $U^*=1/St_n$, where $St_n=f_nD/U_{\infty}$ represents the Strouhal number based on $f_n$. Therefore, with an increase in $U^*$, the stiffness of the oscillation-controlling spring decreases.

For both cases, a structured orthogonal curvilinear $O$-grid is used. The simulations are performed in the moving curvilinear domain where the coordinate system is rigidly attached and oscillates with the body. The corresponding governing compressible Navier-Stokes equations (NSE); boundary conditions; nature of the sponge layer to absorb outgoing acoustic, vortical, and entropic waves and prevent their spurious non-physical reflections from the outer boundary; simulation methodology; and validation with stationary and moving cases are described in detail in \citet{deepak2026}. Simulations are performed with high-accuracy numerical methods for spatial discretization of convective and diffusive derivatives \cite{deepak2026}. Temporal integration is performed by the $4^{\text{th}}$-order accurate Runge-Kutta method, where the coupled structural dynamic equations are simultaneously integrated with the governing compressible NSE \cite{deepak2026}. For the TVIV of the circular cylinder, most of the results are presented for a smaller domain size of $30D$ along the radial direction. For the VIRO of the elliptic cylinder, most of the simulation results are presented for a domain size of $60D$, as also reported in detail in \citet{deepak2026} for $M_{\infty} = 0.2$ and $0.6$. The above smaller-domain results are also validated here with respect to a larger domain size of $1400D$ with identical wall resolution and number of points along the azimuthal direction, for both TVIV and VIRO cases in the vital regions of the response at the lowest and highest Mach numbers. Comparative results of domain size dependence in terms of the amplitude of fluctuation response, aerodynamic coefficients and response frequency are presented in Figs.~\ref{fig_1}--\ref{fig_4}, which show that both response dynamics and aerodynamic coefficients have a close match for both domain sizes, indicating that hydrodynamics are not affected by the domain size. Validation of the results with respect to time step $\Delta t$ and wall resolution is presented in \citet{deepak2026} and is not repeated here.

\section{Results and Discussions}\label{sec3}

     \begin{table}[!ht]   
     \centering
     \resizebox{\textwidth}{!}
     {\begin{tabular}{|c|c|c|c|c|c|c|c|c|c|}
     	 \hline
         & & & & & & & & &             \\                         
     	Geometry & $M_{\infty}$ & $Re$ & $\bar{C}_{l}$ & $({C'_{l}})_{max}$ & $\bar{C}_{d}$ & $({C'_{d}})_{max}$
     	& $\bar{C}_{m}$ & $({C'_{m}})_{max}$ & $St_{s}$ \\
     	\hline 
     	\multirow{3}{*}{Circle} & $0.1$ & $150$  & $0$ & $5.071\times 10^{-1}$ & $1.3059$ & $2.399\times 10^{-2}$  & $-5.378 \times 10^{-6}$  & $5.159\times 10^{-3}$ & $0.185$ \\ \cline{2-10} 
     	& $0.3$ & $150$  & $0$ & $5.383\times 10^{-1}$ & $1.3799$ & $3.301\times 10^{-2}$ & $6.001 \times 10^{-7}$ & $5.697\times 10^{-3}$ & $0.183$ \\ \cline{2-10} 
     	& $0.5$ & $150$  & $0$ & $5.641\times 10^{-1}$ & $1.5519$ & $2.959\times 10^{-2}$ & $-1.207 \times 10^{-5}$ & $9.926\times 10^{-3}$ & $0.182$ \\ \hline
     	
     	\multirow{3}{*}{Ellipse} & $0.2$ & $200$  & $0$ & $3.387\times 10^{-2}$  & $0.5155$ & $9.154\times 10^{-5}$ & $2.677 \times 10^{-6}$ & $2.826\times 10^{-3}$ & $0.2875$ \\ \cline{2-10} 
     	& $0.4$ & $200$  & $0$ & $3.719\times 10^{-2}$  & $0.5505$ & $1.064\times 10^{-4}$ & $-8.138 \times 10^{-8}$ & $3.855\times 10^{-3}$ & $0.2803$ \\ \cline{2-10} 
     	& $0.6$ & $200$  & $0$  & $3.746\times 10^{-2}$ & $0.6221$ & $1.887\times 10^{-4}$ & $-3.111\times 10^{-6}$ & $5.387\times 10^{-3}$ & $0.268$    \\ \hline     	
     \end{tabular}}
     \caption{Mean and fluctuating aerodynamic coefficients of stationary circular and elliptic cylinders are tabulated. Here $({C'_l})_{max}$, $({C'_d})_{max}$ and $({C'_m})_{max}$ represents the amplitude of fluctuation for lift, drag and moment coefficients, respectively, while $\bar{C}_l$, $\bar{C}_d$ and $\bar{C}_m$ represent corresponding time averaged coefficients, respectively}  
     \label{table_1}
     \end{table}
     
\begin{figure}[htbp!]
	\centering
	\includegraphics[width=\textwidth]{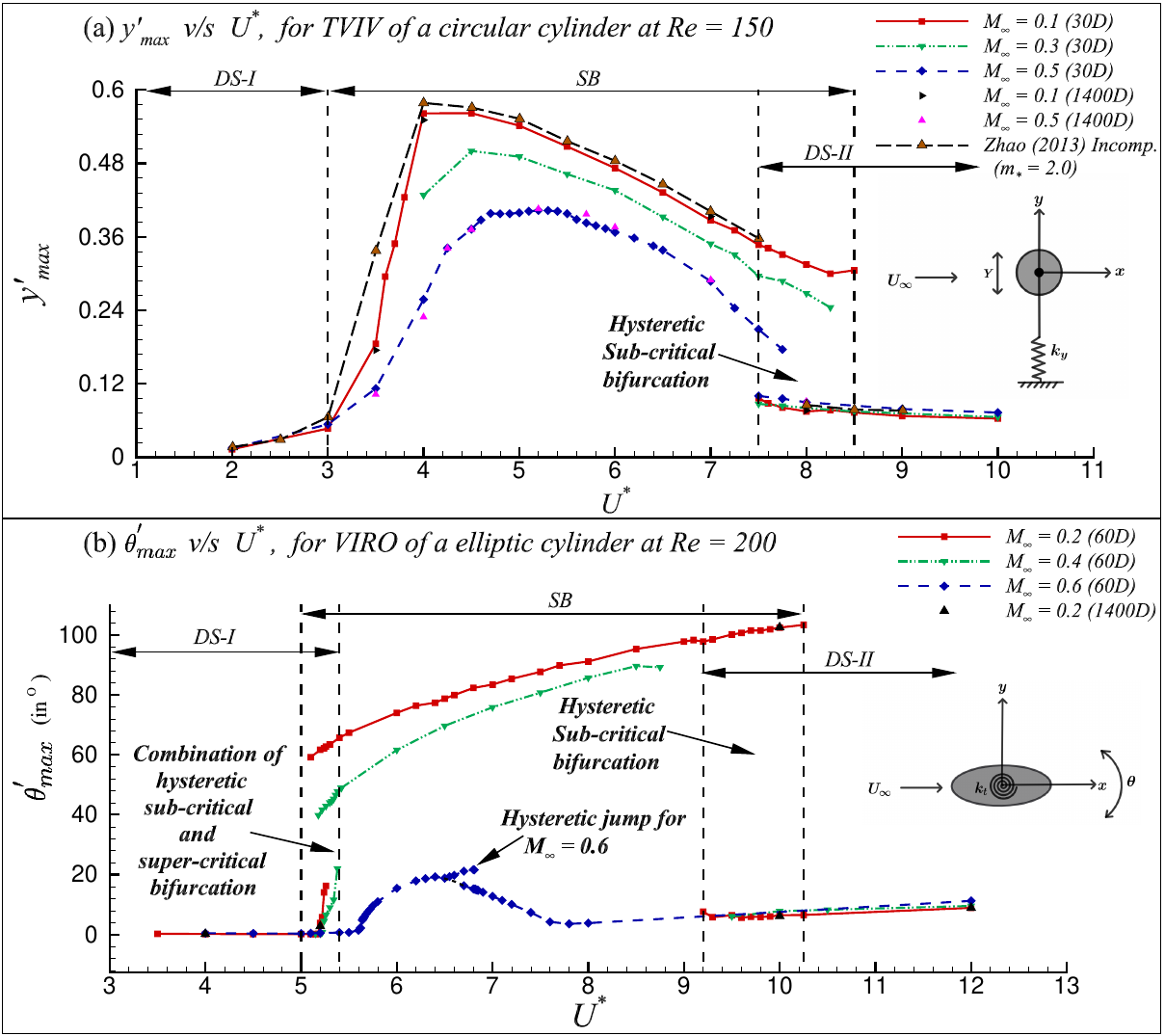} 
	\caption{Variation of oscillation amplitude (a) $y'_{max}$ for TVIV of a circular cylinder ($Re=150$ and $m_*=3.183$) and (b) $\theta'_{max}$ for VIRO of an elliptic cylinder ($A_r=0.5$, $Re=200$, and $m_*=6.366$) plotted as a function of $U^*$.}
	\label{fig_1}
\end{figure}

\begin{figure}[htbp!]
	\centering
	\includegraphics[width=\textwidth]{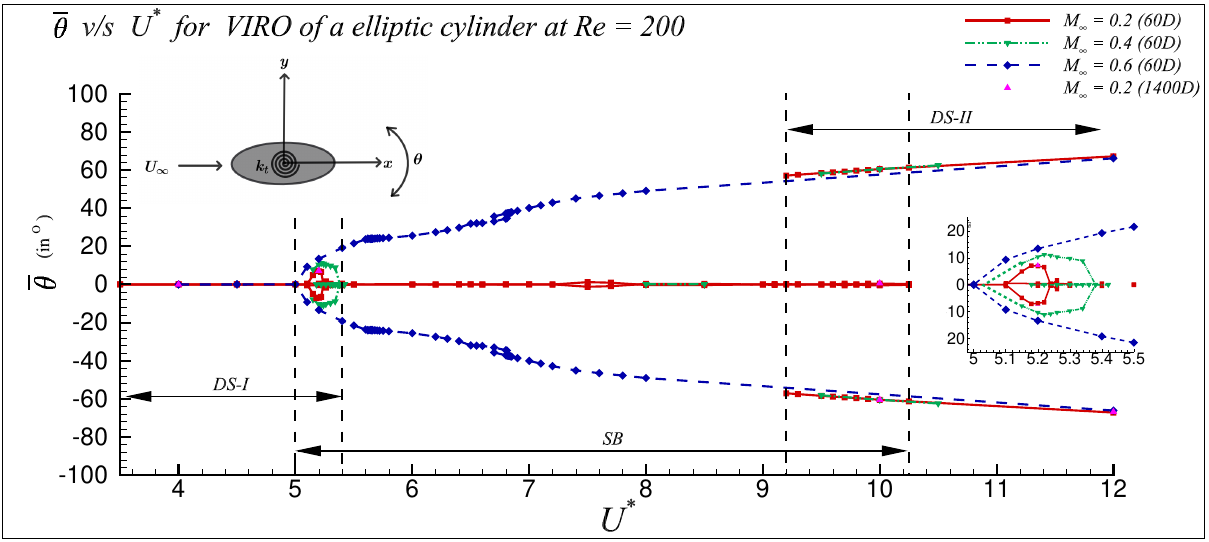}
	\caption{Variation of mean deflection $\bar{\theta}$ from the horizontal orientation plotted as a function of $U^*$ for VIRO of an elliptic cylinder ($A_r=0.5$, $Re=200$, and $m_*=6.366$).}
	\label{fig_2}
\end{figure}

\begin{figure}[htbp!]
	\centering
	\includegraphics[width=\textwidth]{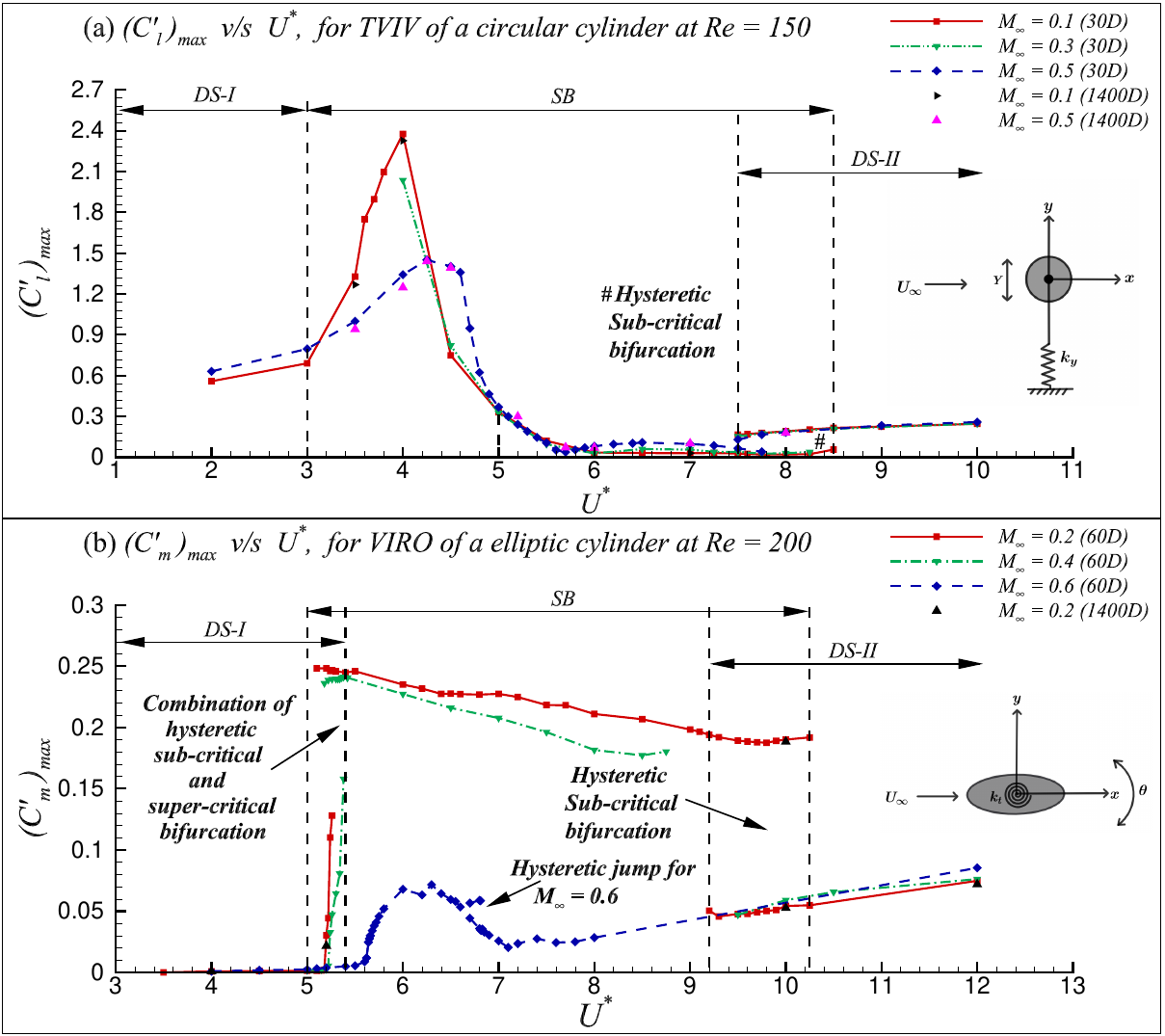}
	\caption{Variation in fluctuation amplitude of (a) lift coefficient $C_{l}$ for TVIV of a circular cylinder ($Re=150$ and $m_*=3.183$) and (b) moment coefficient $C_{m}$ for VIRO of an elliptic cylinder ($A_r=0.5$, $Re=200$, and $m_*=6.366$) plotted as a function of $U^*$.}
	\label{fig_3}
\end{figure}

\begin{figure}[htbp!]
	\centering
	\includegraphics[width=\textwidth]{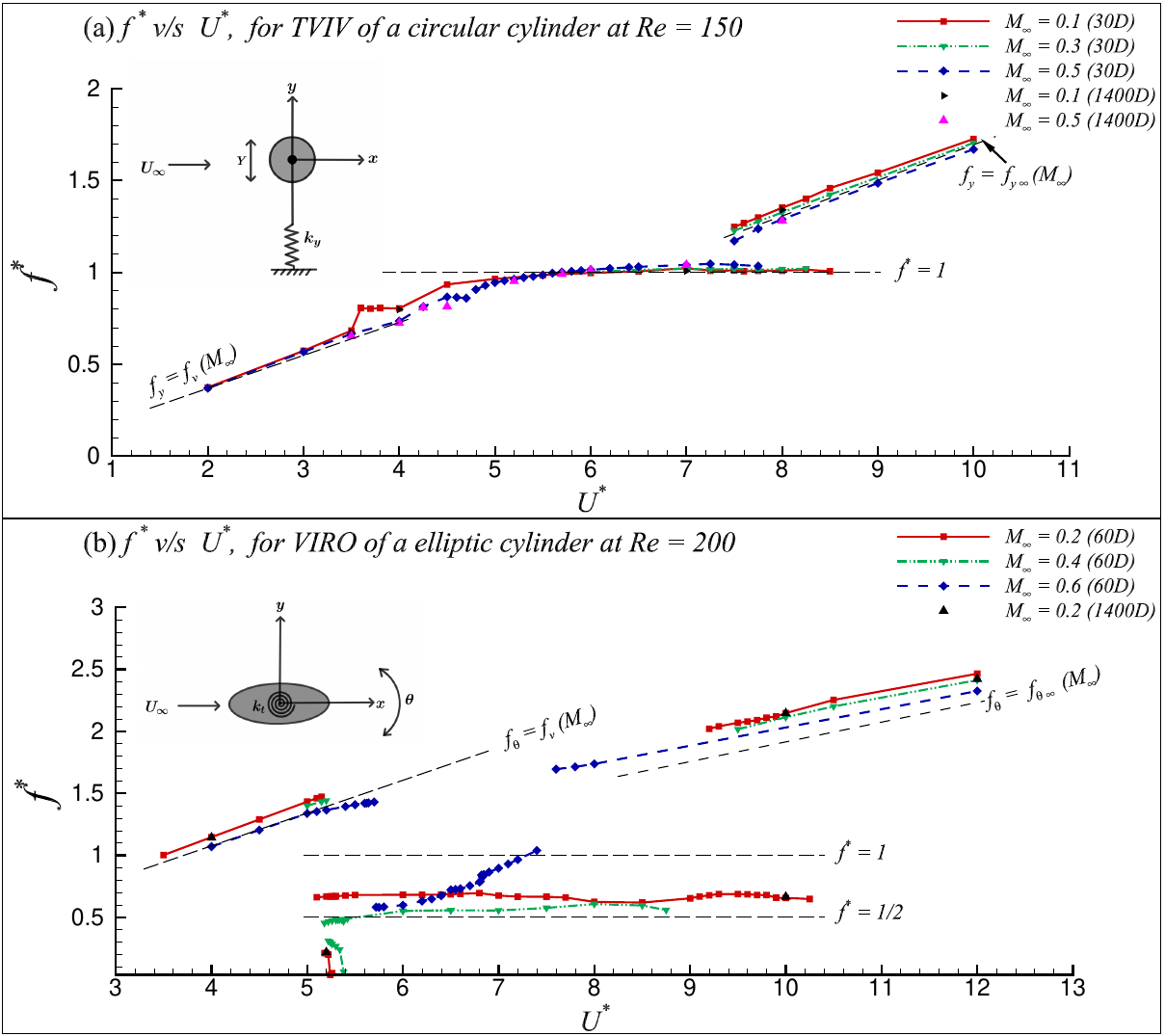}
	\caption{Frequency ratio $f^*$ plotted as a function of $U^*$. Frames~(a) and (b) correspond to TVIV of a circular cylinder ($Re=150$ and $m_*=3.183$) and VIRO of an elliptic cylinder ($A_r=0.5$, $Re=200$, and $m_*=6.366$), respectively.}
	\label{fig_4}
\end{figure}

\begin{figure}[htbp!]
	\centering
	\includegraphics[width=\textwidth]{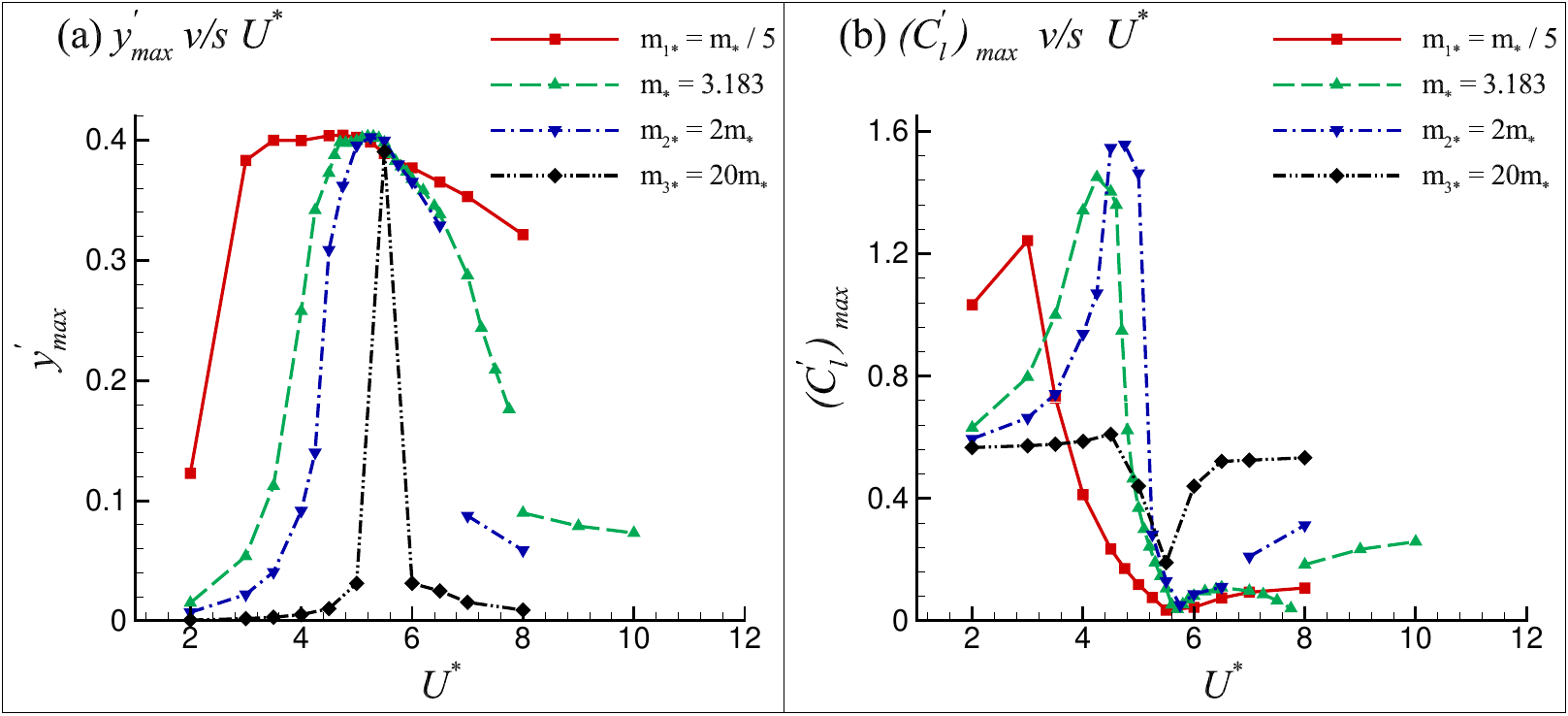}
	\caption{Variation of (a) vibration amplitude $y'_{max}$ and (b) fluctuation amplitude of lift coefficient $C_{l}$ for TVIV of a circular cylinder corresponding to the indicated $m_*$ values plotted as a function of $U^*$ for $M_{\infty}=0.5$ and $Re=150$.}
	\label{fig_5a}
\end{figure}

For both the undamped TVIV of a circular cylinder and the VIRO of an elliptic cylinder of aspect ratio $A_r=0.5$, we take a low $Re$ of $150$ and $200$, respectively, so that the two-dimensional simulations are physically valid. To ascertain the compressibility effects, extensive simulations of undamped TVIV of a circular cylinder are performed for $M_{\infty}=0.1$, $0.3$, and $0.5$ while varying $U^*$. Similar detailed simulations for the VIRO of the elliptic cylinder with $A_r=0.5$ are performed for $M_{\infty}=0.2$, $0.4$, and $0.6$. The corresponding mean and fluctuating aerodynamic coefficients of the stationary circular and elliptic cylinders are tabulated in Table~\ref{table_1}. In the second desynchronous zone $DS\text{-}II$, additional cases of $M_{\infty}$ for the TVIV circular cylinder ($M_{\infty} = 0.2$, $0.4$, and $0.6$) and for the VIRO elliptic cylinder ($M_{\infty}=0.1$, $0.3$, and $0.5$) at $U^*=10$ and $12$, respectively, are simulated. For all cases, results were examined once the statistical stationarity of the mean and fluctuating quantities corresponding to the oscillating cylinder was achieved.

For the stationary circular and elliptic cylinders, the fluctuation amplitudes of aerodynamic coefficients do not exhibit an appreciable increase with an increase in $M_{\infty}$ \cite{canuto2015two}, as noted from Table~\ref{table_1}. The time-averaged values of lift and moment are nearly zero ($(\bar{C}_l,\bar{C}_m)\approx 0$) due to the symmetry of the body with respect to the incoming free-stream flow, while the time-averaged drag coefficient ($\bar{C}_d$) increases with an increase in $M_{\infty}$. It is shown in \citet{canuto2015two} that the increase in $\bar{C}_d$ with $M_{\infty}$ can be satisfactorily predicted by the Prandtl-Glauert (PG) relationship for $Re \ge 100$, though the latter is essentially obtained by relating lift coefficients for inviscid flows. With an increase in $M_{\infty}$, the vortex-shedding Strouhal number $St_{s}$ decreases, as noted in \citet{canuto2015two}, which showed for a stationary flow past a circular cylinder that $St_{s}\approx (-\eta_1 M_{\infty}^2+\eta_0)Re+(\delta_1M_{\infty}^2+\delta_0)$, where the constants $\eta_1$, $\eta_0$, $\delta_1$, and $\delta_0$ are all positive.

For flows over a circular cylinder, it has been demonstrated theoretically, experimentally, and analytically that an imposed transverse oscillation at moderate amplitudes, or vortex-induced vibration (VIV) in the transverse direction, delays the onset of three-dimensional ($3D$) instabilities \cite{leontini2007three,verma2022three,berger1967suppression,koopmann1967vortex,griffin1971unsteady}. Investigations by Floquet stability analysis of a transversely oscillating circular cylinder \cite{leontini2007three} and subsequent numerical investigations \cite{verma2022three} for TVIV indicate that the critical Reynolds number for the onset of $3D$ instabilities for incompressible flow past a circular cylinder can be pushed to as high as $Re_{\text{cr}} = 250\text{--}280$ from $Re_{\text{cr}}=180\text{--}190$ for the stationary case, including cases when the vibration amplitude is as high as $0.6D$ at near lock-in conditions (frequency close to the natural shedding frequency) for both wake modes $A$ and $B$ (see \citet{williamson1988vortex}). For incompressible flows over stationary elliptic cylinders with $A_r=0.5$, the onset of $3D$ instabilities occurs at a higher $Re_{\text{cr}} \approx 250$ \cite{leontini2015stability} because the body is more streamlined than the circular cylinder. The latter can also be pushed higher due to transverse vibration, rotary oscillation \cite{jacono2010modification}, and compressibility effects \cite{canuto2015two}. The rotary oscillation cases further invoke the Taylor--Proudman theorem \cite{tritton2012physical}, which predicts enhanced two-dimensional effects due to rotationality. Therefore, the present two-dimensional simulations at low Reynolds numbers ensure physical validity while simultaneously permitting a broad parametric sweep. Though combining a low $Re$ with higher $M_{\infty}$ lacks real-world physical realizability, the motive behind the current computational framework is to cleanly isolate the influence of compressibility on distinct response branches.

Figures~\ref{fig_1}--\ref{fig_3} describe the compressibility effect on the response of the cylinder for both TVIV of a circular cylinder and VIRO of an elliptic cylinder. Here, $y_{max}^{'}$ is the maximum fluctuation in transverse displacement of the circular cylinder with respect to the mean, $\theta'_{max}=\max(\theta-\bar{\theta})$ is the maximum fluctuation in angular rotation, and $\bar{\theta}$ is the mean angular rotation of the major axis. For both VIV of an elliptic and circular cylinder, the spring is activated once the flow over the corresponding stationary cylinder attains a statistically stationary stage, resulting in flow-induced oscillation.

For both the TVIV case of a circular cylinder and the VIRO case of an elliptic cylinder, the variation in flow dynamics is divided into three prominent branches of response, \textit{viz.} the first desynchronization zone or $DS\text{-}I$; the lock-in zone or the \textit{synchronous branch} or $SB$; and the second desynchronization zone or $DS\text{-}II$ (see Fig.~\ref{fig_1}). While $DS\text{-}I$ is noted before the onset of lock-in, $DS\text{-}II$ exists at the end of lock-in. The $DS\text{-}I$ is noted for small values of $U^*$, while $DS\text{-}II$ exists for large values of $U^*$, \textit{i.e.}, for very stiff and flexible springs, respectively. Similar branchings for TVIV of circular and elliptic cylinders subject to incompressible flow are also highlighted in \citet{mittal2017new,kumar2018identification,navrose2014free,sourav2020pof}, respectively. Figure~\ref{fig_1}(a) also presents the results from \citet{zhao2013flow} corresponding to TVIV of a circular cylinder of $m_*=2.0$ subjected to uniform incompressible flows at $Re=150$. For both the TVIV and VIRO cases presented here, the nature of the flow dynamics is time-harmonic for both desynchronous branches ($DS\text{-}I$ and $DS\text{-}II$). However, for the synchronous branch, intermittent alteration of flow dynamics between a quasi-periodic and periodic nature is observed. Such intermittency between quasi-periodic and periodic flow dynamics in $SB$ for the VIRO case of an elliptic cylinder is illustrated in \citet{deepak2026}. Several interesting flow phenomena are evident in $SB$, including shape reversal and period-doubling bifurcations along with asymmetric vorticity patterns in the wake ($P+S$, $2P+S$, $T+S$, etc., as illustrated in \citet{deepak2026} for VIRO of an elliptic cylinder).

For all $M_{\infty}$ cases of both circular and elliptic cylinders, maximum oscillation amplitudes are noted in the $SB$ zone, while in the $DS\text{-}II$ and $DS\text{-}I$ branches, low oscillation amplitudes are encountered for both TVIV and VIRO cases (refer to Fig.~\ref{fig_1}(a,b)). Fig.~\ref{fig_4} depicts the variation of the frequency ratio $f^*$ with $U^*$ for TVIV and VIRO cases. The frequency ratio $f^*$ is defined as $f^* = f/f_{n}$, where $f$ represents the dominant frequency corresponding to vibration or oscillation motion. In the $DS\text{-}I$ zone, the oscillations of the cylinder for both the TVIV case and the VIRO case are periodic at low $Re$ cases with the frequency of oscillation determined by the vortex shedding frequency (\textit{i.e.}, $f = f_v(M_{\infty})$), as noted in Fig.~\ref{fig_4}. Here, $f_v(M_{\infty})= U_{\infty} St_{s} / D$ represents the vortex shedding frequency of the corresponding stationary cylinder, which is also tabulated in Table~\ref{table_1}. In $DS\text{-}I$, for all the above low $Re$ TVIV and VIRO cases investigated here, a $2S$ vorticity pattern is noted.

For all cases of TVIV, the onset of $SB$ occurs approximately at $U^*\simeq 3.0$ (refer to Fig.~\ref{fig_1}(a)). This indicates that the onset location of $SB$ generally does not depend significantly on $M_{\infty}$ for transverse vibration of a circular cylinder for a fixed $m_{*}$. For transverse vibration, the investigation into the effect of $m_{*}$ indicates the onset is delayed with an increase in $m_{*}$ while keeping the maximum vibration amplitude $y'_{max}$ almost identical (see Fig.~\ref{fig_5a}). For this case, initially there is an increase in response from the end of $DS\text{-}I$ until maximum oscillations are achieved. Figure~\ref{fig_1}(a) indicates that the increase in $y'_{max}$ is steeper for lower $M_{\infty}$. There may be the possibility of a hysteretic jump for the lower Mach number case, which is not further explored here. A similar steeper variation in $y'_{max}$ just before attaining the maximum can also be noted for the higher $M_{\infty}$ case if $m_{*}$ is either very low or very high, as noted in Fig.~\ref{fig_5a}(a).

For VIRO of a symmetrical elliptic cylinder, the onset of $SB$ is also coupled with a supercritical pitchfork bifurcation in mean deflection ($\bar{\theta}$), indicating instability of the mean horizontal orientation ($\bar{\theta} = 0$). The onset of this bifurcation marginally reduces with an increase in $M_{\infty}$, as noted in the inset of Fig.~\ref{fig_2}. Here, the onset of $SB$ for low $M_{\infty}$ cases is associated with the existence of three stable branches (two solution branches for $\theta'_{max}$ and three branches for $\bar{\theta}$) that essentially represent the \textit{hysteretic behavior} of the flow. The corresponding bifurcation has characteristics corresponding to both supercritical and \textit{subcritical} pitchfork bifurcations. While the supercritical bifurcation corresponds to two solutions of $|\bar{\theta}|>0$ with identical $\theta'_{max}$, the subcritical one is associated with a hysteretic jump in $\theta'_{max}$ with $\bar{\theta}\approx 0^{\circ}$. Intermittent switching between two stable branches of the supercritical bifurcation (positive and negative $\bar{\theta}$ attractors) is also noted here for these $M_{\infty}$ cases. Such switching becomes more erratic and intermittent as $U^*$ increases, and the dynamics mimic the \textit{Lorenz system}~\cite{drazin1992}, as highlighted in Fig.~\ref{fig_1}(b). The intensity of such situations is noted to decrease with an increase in $M_{\infty}$, as depicted in Fig.~\ref{fig_1}(b) and Fig.~\ref{fig_2} for $M_{\infty}= 0.2$ and $0.4$, and is completely absent for $M_{\infty}=0.6$. Details in this regard, including the spectral content, time histories, Lissajous plots, and the wake-pattern topology, are demonstrated in \citet{deepak2026}.

The above situations for TVIV and VIRO resemble the primary hysteresis at the beginning of the lock-in zone for TVIV of a symmetric elliptic cylinder subject to incompressible flows\cite{navrose2014free}, where the intensity of the hysteretic jump is noted to depend on the aspect ratio of the cylinder. Similarly, the results here indicate that its intensity also depends on $M_{\infty}$. With an increase in $M_{\infty}$, the intensity of the primary hysteretic jump reduces, and at a higher Mach number, the initial variation of oscillation amplitude becomes smoother. This aspect of the dynamics becomes clear by comparing the corresponding $y'_{max}$ for TVIV of a circular cylinder and $\theta'_{max}$ for the VIRO of an elliptic cylinder. For the TVIV case of a circular cylinder, $y'_{max}$, after achieving maximum amplitude in $SB$, starts to decrease monotonically in $SB$ (see Fig.~\ref{fig_1}(a)). This variation also becomes steeper for relatively higher $m_{*}$, as noted in Fig.~\ref{fig_5a}(a). In contrast, for VIRO, $\theta'_{max}$ increases monotonically in $SB$ for lower $M_{\infty}$ cases from onset to the end. With an increase in $M_{\infty}$, the rate of increase is moderated. For higher $M_{\infty}$ cases like $0.6$, the variation is similar to the TVIV cases, with a monotonic increase up to maximum $\theta'_{max}$ and a decrease thereafter (see Fig.~\ref{fig_1}(b)).

In Fig.~\ref{fig_3}, $C'_{l,\max}$ for TVIV and $C'_{m,\max}$ for VIRO are presented as the respective aerodynamic drivers of their corresponding motions. For TVIV, the variation of $C'_{l,\max}$ is similar to $y'_{max}$, which initially increases with $U^*$ in $SB$, followed by a monotonic decrease (refer to Fig.~\ref{fig_3}(a)). With an increase in $M_{\infty}$, the variation of $C'_{l,\max}$ becomes smoother and less steep. In contrast, for VIRO at lower $M_{\infty}$, the variation of $C'_{m,\max}$ in $SB$ decreases monotonically, in contrast to the monotonic increase of $\theta'_{max}$ in this zone. However, this variation is altered at higher $M_{\infty}$, where the variation is similar to the corresponding variation of $\theta'_{max}$ (see Fig.~\ref{fig_3}(b)).

In the $SB$ zone, the value of $f^*\simeq 1.0$ for TVIV of a circular cylinder for all $M_{\infty}$ cases (see Fig.~\ref{fig_4}(a)), indicating a resonance phenomenon associated with a high oscillation amplitude (see Fig.~\ref{fig_1}(a)). Similar observations were also noted for TVIV of a circular cylinder \cite{leontini2007three,sahu2019flow,verma2023three}, a diamond cylinder \cite{sourav2020undamped}, a triangular cylinder \cite{sourav2023deciphering}, and an elliptic cylinder with varying angle of attack \cite{zhao2019dynamic,leontini2018,yadav2024flow} subjected to incompressible flows with $Re$ varying from low ($Re=100\text{--}150$) to high values ($Re=1000$) corresponding to $2D$ and $3D$ flow dynamics. The dynamics for VIRO are significantly different from the TVIV case in $SB$ (see Fig.~\ref{fig_4}(b)), where for lower $M_{\infty}$ cases, $f^*$ in $SB$ is almost invariant with $U^*$ such that $f^* \approx 0.5$ in this zone. For higher $M_{\infty}$ cases, $f^*$ gradually increases from $0.5$ to $1.0$, as noted for $M_{\infty} = 0.6$ in Fig.~\ref{fig_4}(b). For this case, the oscillation frequency is slightly less than $f_v/2$ in $SB$. Hence, while for low $M_{\infty}$ cases, the cylinder oscillates almost at $f_n/2$, it tends to oscillate almost at $f_v/2$ for higher $M_{\infty}$ cases in $SB$. This represents a gradual transition of the \textit{period-doubling phenomenon} \cite{leontini2018,yadav2024flow} from half of the natural structural frequency to half of the vortex-shedding frequency of the corresponding stationary cylinder for the VIRO case.

For both TVIV and VIRO cases, the secondary hysteretic jump at the end of $SB$ is noted, similar to that reported in \citet{navrose2014free} for TVIV of an elliptic cylinder subject to incompressible flows. The extent and intensity of this secondary hysteretic jump reduce with $M_{\infty}$ for both TVIV and VIRO, as noted in Figs.~\ref{fig_1}--\ref{fig_3}. For example, for TVIV, the secondary hysteretic zone is noted for $7.5 \leq U^* \leq 8.5$ for $M_{\infty} = 0.1$, $7.5 \leq U^* \leq 8.2$ for $M_{\infty} = 0.3$, and $7.5 \leq U^* \leq 7.75$ for $M_{\infty} = 0.5$. For the higher $M_{\infty} = 0.6$ case of VIRO, the secondary hysteretic jump completely disappears at the end of the lock-in zone, though a new intermediate hysteretic jump between a quasi-periodic state and a periodic state of orbit-1 is noted close to the $U^*$ where the oscillation amplitude is maximum (see Fig.~\ref{fig_1}(b)). More details regarding the dynamics of VIRO for $M_{\infty} = 0.2$ and $0.6$ are given in \citet{deepak2026}.

The lower branch of the hysteretic jump further continues to $DS\text{-}II$ (see red, green, and blue lines for $M_{\infty} = 0.1$, $0.3$, and $0.5$, respectively, for TVIV, and red and green lines for $M_{\infty} = 0.2$ and $0.4$, respectively, for VIRO), while for $M_{\infty}=0.6$, the variation smoothly merges with $DS\text{-}II$. For VIRO, even though there is a sudden drop in the oscillation amplitude from the end of the lock-in zone to the onset of $DS\text{-}II$, $\theta'_{max}$ and $C'_{m,\max}$ continue to increase monotonically in this regime until the corresponding free-to-rotate condition ($U^*\rightarrow \infty$). In this regime, a $C(2S)$ pattern of the vortex street is noted irrespective of $M_{\infty}$, and the mean deflection $\bar{\theta}$ continues to increase, indicating that the effective angle of attack also increases with $U^*$. In contrast, for TVIV, the oscillation amplitude $y'_{max}$ monotonically decreases while $C'_{l,\max}$ increases in $DS\text{-}II$. Similar observations are also reported for undamped TVIV of a circular cylinder subjected to incompressible flow in \citet{soti2020vortex} (for $Re = 150$, $100$ and $m_* = 10$), \citet{zhao2020effects} (for $Re = 150$ and $m_* = 8.0/\pi$), and \citet{prasanth2008vortex} (for $Re = 100$ and $m_* = 10.0$). The apparent reason for such behavior may be explained by considering the relative dominance of higher harmonics for the periodic motion noted in $DS\text{-}II$ for TVIV. For VIRO, even though similar spectral behavior is also noted in $DS\text{-}II$ \cite{deepak2026}, $\theta'_{max}$ increases due to an increase in the effective angle of attack, leading to oscillation characteristics that resemble galloping.

Figure~\ref{fig_4} also shows that for both TVIV and VIRO, the frequency ratio $f^*$ linearly increases with $U^*$ in $DS\text{-}II$ as the cylinder oscillates with a frequency that is close to the corresponding frequency of free transverse vibration ($f_{y\infty}$) or oscillation ($f_{\theta\infty}$). Figure~\ref{fig_4} indicates that the free-to-oscillate frequency increases marginally with a decrease in $M_{\infty}$, similar to the vortex-shedding behavior of the corresponding stationary cylinder as elaborated in Table~\ref{table_1}. In fact, for TVIV of a circular cylinder, $f_{y\infty} \simeq 0.93 f_v$, while for VIRO of the elliptic cylinder, $f_{\theta\infty} \simeq 0.71 f_v$, independent of $M_{\infty}$, where $f_v$ represents the vortex-shedding frequency of the corresponding stationary cylinder, as noted in Table~\ref{table_1}.

\begin{figure}[htbp!]
	\centering
	\includegraphics[width=\textwidth]{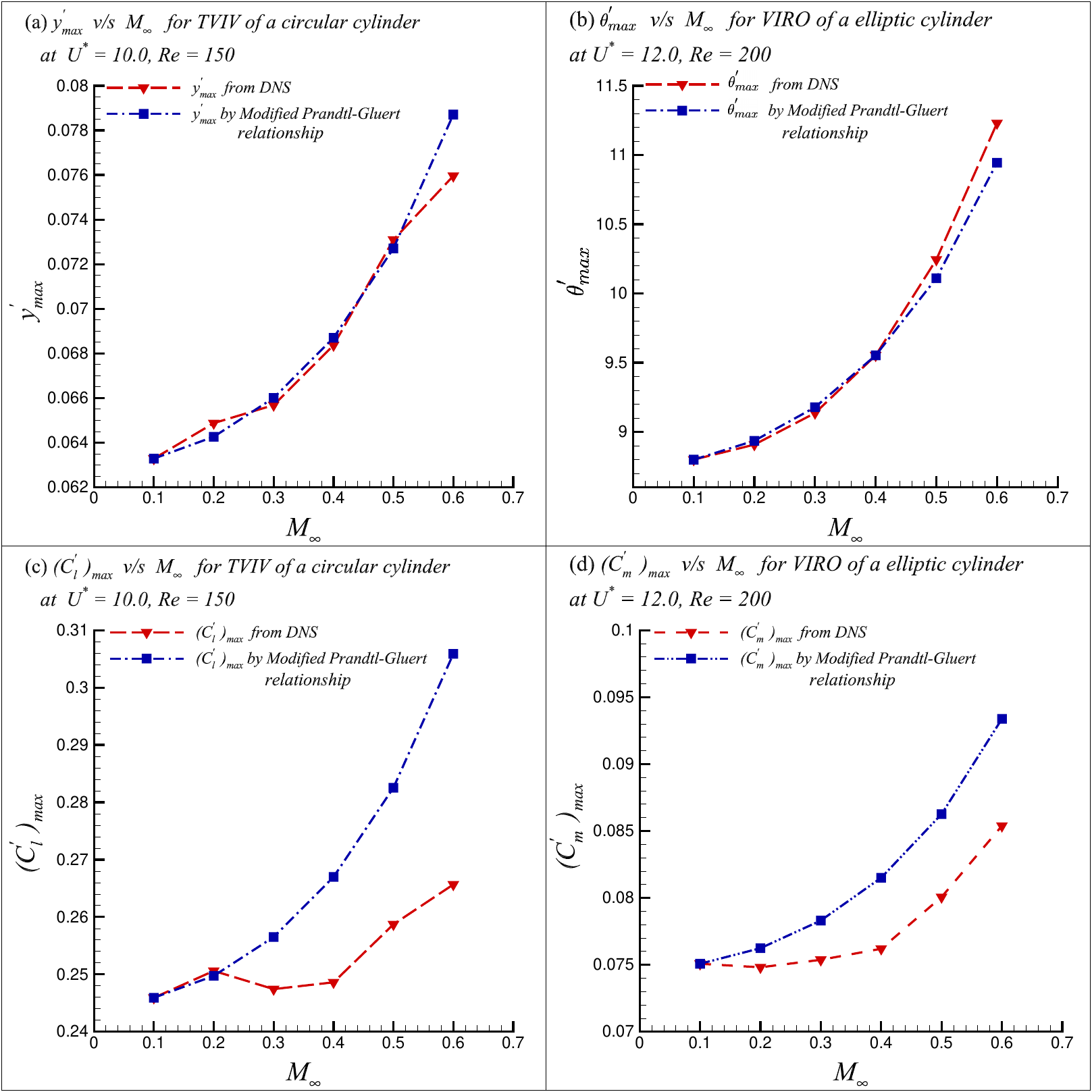}
	\caption{Variation of oscillation amplitude ($y'_{max}$ and $\theta'_{max}$) and fluctuation amplitude of lift and moment coefficients ($C'_{l,\max}$ and $C'_{m,\max}$) for TVIV of a circular cylinder at $U^*=10$ in (a,c) and VIRO of an elliptic cylinder at $U^*=12$ in (b,d) plotted as a function of $M_{\infty}$ along with the \textit{modified Prandtl--Glauert relationship}.}
	\label{fig_5}
\end{figure}

In $SB$, the results presented through Figs.~\ref{fig_1}--\ref{fig_3} indicate enhanced stabilization of the flow-induced motion due to an increase in $M_{\infty}$ \cite{canuto2015two}, leading to a decrease in oscillation amplitude for both TVIV and VIRO. In contrast, in $DS\text{-}II$, the oscillation amplitudes for both TVIV and VIRO increase with an increase in $M_{\infty}$. A similar increase is also noted with respect to $C'_{l,\max}$ and $C'_{m,\max}$ for TVIV and VIRO, respectively. This phenomenon mimics the Prandtl-Glauert (PG) compressibility correction analogy \cite{anderson2017}, which is derived based on inviscid and irrotational flow assumptions. The above rule postulates the increase in aerodynamic coefficients with an increase in $M_{\infty}$, which is given as

\begin{eqnarray}
  C_{l} = \frac{C_{l_i}}{\sqrt{1-M_{\infty}^2}} \quad \text{and} \quad
  C_{m} = \frac{C_{m_i}}{\sqrt{1-M_{\infty}^2}}
\end{eqnarray}

\noindent where $C_{l_i}$ and $C_{m_i}$ represent the lift and moment coefficients for the corresponding incompressible flow regime, respectively. Figure~\ref{fig_5} shows a comparison of $y'_{max}$ and $C'_{l,\max}$ for TVIV and $\theta'_{max}$ and $C'_{m,\max}$ for VIRO as obtained from DNS with the \textit{Prandtl-Glauert correlation} for $U^* = 10.0$ and $12.0$, respectively. The corresponding modified PG correlation factor is calculated as
\begin{eqnarray}
  f_{\text{PG}}=\frac{\sqrt{1-(M_{\infty})_{\text{ref}}^2}}{\sqrt{1-(M_{\infty})^2}}
\end{eqnarray}
\noindent where $(M_{\infty})_{\text{ref}}=0.1$ is the reference Mach number. Figure~\ref{fig_5}(a) compares $y'_{max}$ with $(y'_{max})_{\text{PG}} = (y'_{max})_{\text{ref}} \times f_{\text{PG}}$, while Fig.~\ref{fig_5}(c) compares $C'_{l,\max}$ with $(C'_{l,\max})_{\text{PG}} = (C'_{l,\max})_{\text{ref}} \times f_{\text{PG}}$ as a function of $M_{\infty}$ for TVIV at $U^* = 10.0$. A similar correlation is also used in Figs.~\ref{fig_5}(b,d) for $\theta'_{max}$ and $C'_{m,\max}$ for VIRO at $U^* = 12.0$. A good match is obtained for $y'_{max}$ and $\theta'_{max}$ with the corresponding PG correlated values, as noted in Figs.~\ref{fig_5}(a,b). This probably indicates the role of the inviscid nature of compressibility in the $DS\text{-}II$ regime in increasing the oscillation response, as opposed to $SB$, where enhanced viscous stabilization with increasing $M_{\infty}$ leads to the moderation of response. A similar phenomenon as in $DS\text{-}II$ may be happening in $DS\text{-}I$ as well, which is not explored here in detail. As galloping phenomena are noted for non-symmetric cross-sections in the $DS\text{-}II$ regime, the results indicate that, unlike in $SB$, an increase in Mach number will enhance the tendency of galloping oscillation. Although the present simulations are performed at low Reynolds numbers, the identified trends are consistent with fundamental compressibility mechanisms \cite{canuto2015two,nagata2020experimental} and may therefore persist at higher Reynolds numbers. Further three-dimensional investigations at higher Reynolds numbers are needed to confirm the quantitative extent of these trends.

 \section{Summary and Conclusions}\label{sec4}

 The comparative study of transverse vortex-induced vibration (TVIV) of a circular cylinder and vortex-induced rotary oscillation (VIRO) of an elliptic cylinder of aspect ratio $0.5$ is conducted for $0.1\leq M_{\infty} \leq 0.6$ cases to gain a broader understanding of the common features of the compressibility effects on flow-induced motion in different characteristic regimes for low to moderate subsonic Mach number cases. For both cases of flow-induced motion, the general existence of three characteristic branches, $viz.,$ $DS\text{-}I$ (Desynchronous zone $I$), $SB$ (synchronous branch), and $DS\text{-}II$ (Desynchronous zone $II$), is noted. Though the dynamics are significantly different for TVIV and VIRO, as noted in Figs.~\ref{fig_1}--\ref{fig_4}, several common features are also observed, such as enhanced oscillation amplitude in the lock-in zone $SB$ and reduced effects in the other two zones. For both cases of flow-induced motion, significant suppression of oscillation amplitude and fluctuating aerodynamic coefficients in $SB$ is noted with an increase in $M_{\infty}$. The increase in $M_{\infty}$ also reduces the extent and intensity of hysteretic jumps at the onset and end of $SB$, as demonstrated in Figs.~\ref{fig_1} and \ref{fig_3}. Despite the stabilizing effects of compressibility in $SB$, the fluctuation amplitudes are noted to increase in $DS\text{-}II$ with $M_{\infty}$ for both cases of flow-induced motion. Interestingly, for both cases, the variation of oscillation amplitude in the $DS\text{-}II$ zone closely follows a \textit{modified Prandtl-Glauert relationship} (see Fig.~\ref{fig_5}). This dichotomy indicates that viscous stabilization dominates in the lock-in regime, while inviscid compressibility effects govern the response in $DS\text{-}II$. Although the present simulations are performed at low Reynolds numbers, the identified trends are consistent with fundamental compressibility mechanisms and may therefore persist at higher Reynolds numbers. It is also worth noting that in physical flows an increase in free-stream Mach number is typically accompanied by a rise in effective Reynolds number (an effect that can be further accentuated in wind-tunnel experiments with fixed stagnation conditions), which may further enhance the amplitude trends observed here in the $DS\text{-}II$ regime. Quantitative confirmation of these effects through three-dimensional simulations at higher Reynolds numbers remains a subject for future investigation.

\bibliography{Reference}
\end{document}